\documentclass[twocolumn,showpacs,floatfix,superscriptaddress,amsmath,amssymb,prl]{revtex4}
\usepackage{bbm}
\usepackage{mathrsfs}
\usepackage{txfonts}
\usepackage{amssymb}
\usepackage{graphicx}
\usepackage{hyperref}
\usepackage{ulem}
 \usepackage{overpic}
 \usepackage{psfrag}
\makeatletter
\renewcommand\subsection{\@startsection
{subsection}{2}{0mm}
 {-\baselineskip}
 {0.5\baselineskip}
{\FloatBarrier\normalfont\Large\bfseries}}
\makeatother
\newcommand{\be}{\begin{equation}}
\newcommand{\ee}{\end{equation}}

\newcommand{\PreserveBackslash}[1]{\let\temp=\\#1\let\\=\temp}

\newcommand{\ket} [1] {| #1 \rangle}

\begin{document}
\title{Geometric entanglement from matrix product state representations}

\author{Bing-Quan Hu, Xi-Jing Liu, Jin-Hua Liu and Huan-Qiang Zhou}
\affiliation{Centre for Modern Physics and Department of Physics,
Chongqing University, Chongqing 400044, The People's Republic of
China}

\begin{abstract}
An efficient scheme to compute the geometric entanglement per
lattice site for quantum many-body systems on a periodic finite-size
chain is proposed in the context of a tensor network algorithm based
on the matrix product state representations. It is systematically
tested for three prototypical critical quantum spin chains, which
belong to the same Ising universality class. The simulation results
lend strong support to the previous claim [Q.-Q. Shi, R. Or\'{u}s,
J. O. Fj{\ae}restad, and H.-Q. Zhou, New J. Phys \textbf{12}, 025008
(2010); J.-M. St\'{e}phan, G. Misguich, and F. Alet, Phys. Rev. B
\textbf{82}, 180406R (2010)] that the leading finite-size correction
to the geometric entanglement per lattice site is universal, with
its remarkable connection to the celebrated Affleck-Ludwig boundary
entropy corresponding to a conformally invariant boundary condition.

\end{abstract}


\maketitle

{\it Introduction.}  In the last decade, significant progress has
been made in the investigation of quantum phase transitions (QPTs)
from a novel perspective-quantum entanglement~\cite{amico}. The main
idea is to quantify entanglement present in a ground-state wave
function for a quantum many-body lattice system in terms of a
variety of bipartite entanglement measures, with some anomaly at a
critical point. A remarkable result is achieved for the von Neumann
entropy that quantifies the bipartite entanglement when a
finite-size spin chain is partitioned into two disjoint parts: it
scales logarithmically with the subsystem's size, with a prefactor
proportional to the central charge, a fundamental quantity in
conformal field theory, as long as the system is at
criticality~\cite{latorre,korepin,calabrese,zbfs}.

Contrary to bipartite entanglement measures, no consensus has been
achieved regarding multipartite entanglement measures, although much
effort has been devoted to characterize quantum criticality from
some multipartite entanglement measures. One promising candidate
among them is the geometric entanglement (GE)~\cite{hba,tcw}. As a
holistic measure of the multipartite entanglement, the GE quantifies
the multipartite entanglement present in a quantum state wave
function. For a quantum many-body lattice system, the GE per lattice
site is shown to be an alternative way to detect critical
points~\cite{tcw,abot,rt,rtt,hl}.

In addition, an intriguing connection between the GE per site and
the Affleck-Ludwig $g$ factor~\cite{affleck} is established for a
finite-size spin chain with the periodic boundary conditions at
criticality. More precisely, the GE per site, $\varepsilon_{N}$,
scales as
\begin{equation}
{\mathcal E}_{N} \sim {\mathcal E}_{\infty}+\frac{b}{N} +
O\left(\frac{1}{N^2}\right),
\end{equation}
with $N$ being the lattice size. It was conjectured that the
coefficient $b$ in the subleading term is universal~\cite{Qian}.
This is confirmed later in Ref. \cite{Misguich} for both the
transverse Ising and XXZ chains at criticality, by relating the
coefficient $b$ to the Affleck-Ludwig boundary entropy $s$:
\begin{equation}\label {bs}
   b = -\frac{2}{\ln\;2} \; s.
 \end{equation}
Here, the boundary entropy $s$ is defined through the Affleck-Ludwig
$g$ factor: $s = \ln\; g$~\cite{latorre,korepin,calabrese,zbfs}.
This result is surprising, in the sense that the Affleck-Ludwig
boundary entropy $s$ appears in the GE per lattice site for a {\it
periodic} chain. However, it still remains unclear as to whether or
not such a connection between the GE per site and the Affleck-Ludwig
$g$ factor is universally valid, and whether or not there is any
criterion to judge which $g$ factor, corresponding to a conformally
invariant boundary condition, is chosen for a specific model.
Therefore, it is desirable to investigate the models belonging to
the same universality class, to see if the same $g$ factor appears,
thus providing further insights into the connection between the
subleading term coefficient $b$ in the GE per lattice site and the
Affleck-Ludwig $g$ factor.

One of the main obstacles to address these issues is due to the fact
that the computation of the GE per lattice site is a formidable task
for a quantum many-body lattice system, because it involves the
optimization over all the possible separable states, with possibly
some constraints arising from the translational invariance. However,
recent progress in the context of the tensor network algorithms for
quantum many-body lattice systems with the periodic boundary
conditions~\cite{twe,pwe,a5,pvv,rgf} offers us an efficient method
to systematically evaluate the GE per lattice site for quantum
many-body lattice systems. In this paper, we first describe how to
evaluate the GE per lattice site for a finite-size quantum spin
chain from a tensor network algorithm based on the matrix product
state representations. Then we present our simulation results for
three prototypical spin chains belonging to the Ising universality
class.

{\it The geometric entanglement per lattice site.} We first
introduce the maximum fidelity $\Lambda_{{\rm max}}$ between a
quantum pure state $|\psi\rangle$ and all the possible separable and
normalized states $|\phi\rangle$ of the $N$ parties:
 \begin{equation}\label{fid}
   \Lambda_{{\rm max} }={\rm \max_{|\phi\rangle }} \; |\langle\phi|\psi\rangle|.
 \end{equation}
The larger $\Lambda_{{\rm max}}$ is, the closer a quantum state wave
function is to a separable state. Therefore, a holistic measure of
the multipartite entanglement present in a quantum state wave
function $|\psi\rangle$ is defined as
 \begin{equation}\label{ge}
   E(|\psi\rangle)=-\log_2{\Lambda_{{\rm max}}^{2}}.
 \end{equation}
Since the contribution to $E(\psi)$ from each party is additive,
$E(\psi)$ scales linearly with  $N$, for a multipartite system
consisting of $N$ parties. Therefore, it is convenient to define the
GE per party as
\begin{equation}\label{geper}
   {\mathcal E}_{N}(|\psi\rangle)=N^{-1}E(|\psi\rangle).
 \end{equation}
For a quantum many-body lattice spin system, each lattice site
constitutes a party. Thus, ${\mathcal E}_{N}$ is the GE per lattice
site. Note that it is well defined even in the thermodynamic limit.

\begin{figure}
\begin{overpic}[width=80mm,totalheight=35mm]{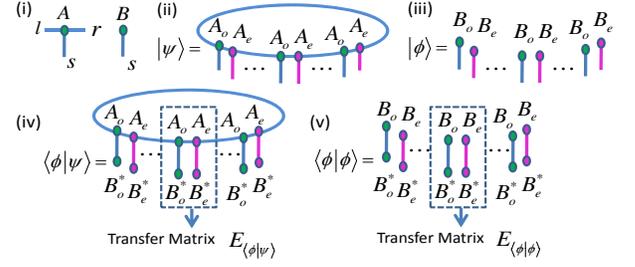}
\end{overpic}
\caption{(color online) (i) Two three-index tensors $A_{o}^{lrs}$,
and $A_{e}^{lrs}$ are attached to odd and even sites respectively,
to represent a ground-state wave function. Two one-index tensors
$B_{o}^{s}$ and $B_{e}^{s}$, attached to odd and even sites
respectively, represent a separable state. Here, $s$ is the physical
index, $l$, and $r$ are the inner bond indices.  (ii) The pictorial
representation for a ground-state wave function $|\psi\rangle$.
(iii) The tensor network representation for a separable state
$|\phi\rangle$. (iv) The fidelity between the ground-state wave
function $|\psi\rangle$ and a separable state $|\phi\rangle$.  The
transfer matrix $E_{\langle\phi|\psi\rangle}$ for the fidelity is
constructed from the tensors $A_{o}$, $A_{e}$, $B_{o}^{*}$ and
$B_{e}^{*}$. (v) The norm for a separable state $|\phi\rangle$,
where the transfer matrix $E_{\langle\phi|\phi\rangle}$ is
constructed from $B_{o}$, $B_{e}$, and their conjugates. }
  \label{draw}
 \end{figure}

{\it A matrix product state algorithm.}  Let us recall the key steps
to produce a ground-state wave function from an efficient
variational algorithm~\cite{a5} for a translational-invariant
finite-size periodic lattice system. First, choose a random state as
an initial state $\ket{\psi_{0}}$. Second, perform the imaginary
time evolution for $\ket{\psi_{0}}$. Thus, we get the evolved state
$\ket{\psi_{\tau}}$ at imaginary time $\tau$,
\begin{equation}
\label{initialstate}
\ket{\psi_{\tau}}=\frac{\exp(-H\tau)\ket{\psi_{0}}}{\|\exp(-H\tau)\ket{\psi_{0}}\|}.
\end{equation}
If $\tau\rightarrow\infty$, the ground-state wave function is
projected out, as long as the initial state is not orthogonal to the
real ground state. Third, exploit the Trotter-Suzuki decomposition
and turn the imaginary time evolution operator into a series of
two-site gates over a time slice $\delta \tau$, with $\tau = M
\delta \tau$. Therefore, the problem becomes how to absorb a
two-site gate acted on a matrix product state.  For a periodic spin
chain with $N$ sites, we assume the matrix product state
representation is translation-invariant under two-site shifts. Thus
we only need two three-index tensors, $A_{o}$ and $A_{e}$ (cf.
Fig.~\ref{draw} (i)), to represent a tensor network for the
ground-state wave function and two one-index tensors, $B_{o}$ and
$B_{e}$ (cf. Fig.~\ref{draw} (i)), to represent a separable state,
as shown in Fig.~\ref{draw} (ii) and (iii), respectively. Here, the
subscripts $o$ and $e$ represent  odd and even sites, respectively.
In this setting, a two-site gate is absorbed  by performing a
singular value decomposition for a matrix contracted from a few
tensors involving the two-site gate.

A peculiar feature of the algorithm for a finite-size periodic
system is to take account of the $p$ largest eigenvalues and the
corresponding eigenvectors of the transfer matrix
$E_{\langle\psi|\psi\rangle}$, when one evaluates the ground-state
energy per site. That is, $ \label{SVDlargem}
E_{\langle\psi\psi\rangle}^{m}=\sum_{i=1}^{D^{2}}u_{i}s_{i}^{m}v_{i}\simeq\sum_{i=1}^{p}u_{i}s_{i}^{m}v_{i},
$ where $s_{i}$ is the $i$-th largest eigenvalue, $u_{i}$ and
$v_{i}$ indicate the left and right eigenvectors, respectively, with
$D$ denoting the bond dimension. When the energy per site converges,
the ground-state wave function is generated. Note that the
computational cost of the algorithm is $p D^{3}$.

\begin{figure}
\begin{overpic}[width=80mm,totalheight=40mm]{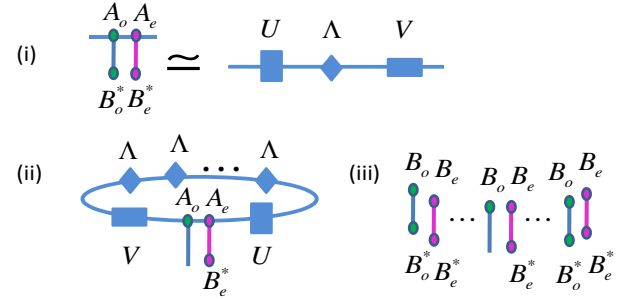}
\end{overpic}
\caption{(color online) (i) The transfer matrix
$E_{\langle\phi|\psi\rangle}$ may be approximated by the $p$ largest
eigenvalues $\Lambda$ and the corresponding left and right
eigenvectors, $U$ and $V$, of the transfer matrix
$E_{\langle\phi|\psi\rangle}$. (ii) The pictorial representation for
the derivative of $\langle\phi|\psi\rangle$ with respect to
$B_{o}^{*}$. (iii) The pictorial representation for the derivative
of $\langle\phi|\phi\rangle $ with respect to $B_{o}^{*}$.}
  \label{draw2}
   \end{figure}

{\it The geometric entanglement per lattice site from matrix product
state representations.} Now we are ready to compute the GE per
lattice site. The crucial step is how to maximize the fidelity
$|\langle\phi|\psi\rangle|$ over all the possible separable states
$|\phi\rangle$. In Fig.~\ref{draw} (iv), we introduce the transfer
matrix $E_{\langle\phi|\psi\rangle}$ to represent the fidelity
between $|\psi\rangle$ and $|\phi\rangle$, with
$E_{\langle\phi|\psi\rangle}$ constructed from two three-index
tensors, $A_{o}$ and $A_{e}$, and two one-index tensors, $B_{o}^{*}$
and $B_{e}^{*}$. Mathematically, the fidelity between a ground-state
wave function $|\psi\rangle$ and a separable state $|\phi\rangle$ is
expressed as:
\begin{equation}
\label{fidelity}
f=\frac{|\langle\phi|\psi\rangle|}{\sqrt{\langle\psi|\psi\rangle
\langle\phi|\phi\rangle}} \,\,.
\end{equation}
For our purpose, we consider $F=f^2$ instead of $f$ itself. To
maximize $F$, we compute its logarithmic gradient with respect to a
one-index tensor $B^{*}$,
\begin{equation}
\label{gradient} \frac{\partial \ln F}{\partial B^{*}}=\frac
{1}{\langle\phi|\psi\rangle} \frac {\partial
\langle\phi|\psi\rangle} {\partial B^{*}}-\frac
{1}{\langle\phi|\phi\rangle} \frac {\partial
\langle\phi|\phi\rangle} {B^{*}} \,\,.
\end{equation}
Here, $B^{*}$ is either $B_{o}^{*}$ or $B_{e}^{*}$.  We take
$B_{o}^{*}$ as an example to explain the updating procedure. The
pictorial representation for the derivative of $\ln F$ with respect
to $B_{o}^{*}$ is shown in Fig.~\ref{draw2}. In Fig.~\ref{draw2}(i),
$E_{\langle\phi|\psi\rangle}$ is approximated by  the first $p$
largest eigenvalues and the corresponding eigenvectors of
$E_{\langle\phi|\psi\rangle}$. That implies that $\label{approxi}
E_{\langle\phi|\psi\rangle}^{m}\simeq \sum_{k=1}^{p}
U_{k}\Lambda_{k,k}^{m}V_{k}$, with $p$ specifically depending on
$m$: the larger $m$ is, the less $p$ is. In Fig.~\ref{draw2}(ii),
the derivative of $\langle\phi|\psi\rangle $ with respect to
$B_{o}^{*}$ is shown. In Fig.~\ref{draw2}(iii), we represent the
derivative of $\langle\phi|\phi\rangle $ with respect to
$B_{o}^{*}$. Once the fidelity gradient is determined, the real and
imaginary parts of $B_{o}$ may be updated as follows:
\begin{subequations}
\begin{equation}
\label{updateRe} \Re B_{o}=\Re B_{o} + \delta \Re (\frac{\partial
F}{\partial B_{o}}),
\end{equation}
and
\begin{equation}
\label{updateIm}
 \Im B_{o}=\Im B_{o} + \delta \Im (\frac{\partial
F}{\partial B_{o}}),
\end{equation}
\end{subequations}
where $\delta$ is the step size in the parameter space, which is
tuned to be decreasing during the updating procedure. Here, we have
normalized the tensors $B_{o}$, $\Re ({\partial F}/{\partial
B_{o}})$ and $\Im ({\partial F}/{\partial B_{o}})$  by setting their
respective largest entries to be unity. In exactly the same way, we
may update the other tensor $B_{e}$. Repeating the procedure until
the fidelity per lattice site converges, we achieve the closest
separable state $|\phi\rangle$ that ensures its maximum fidelity
with the ground-state wave function $|\psi\rangle$. Thus, the GE per
lattice site follows, as desired.

{\it The models.} We test our scheme by considering three
prototypical critical quantum spin chains with the periodic boundary
conditions. Note that the models belong to the same Ising
universality class.

The first model is quantum Ising model in a transverse magnetic
field on a finite-size ring. The Hamiltonian takes the form:
\begin{equation}\label {fitIsing}
   H = -\sum_{i=1}^{N} \left(\sigma_{x}^{[i]} \sigma_{x}^{[i+1]} +  \lambda \sigma_{z}^{[i]}\right),
 \end{equation}
where $\sigma_{\alpha}^{[i]} \; (\alpha = x,z)$ are the Pauli spin
operators at site $i$, and $\lambda$ is a transverse magnetic field.
The model is critical at $\lambda_c =1$.

The second model is a spin-1/2 Ising chain with antisymmetric
anisotropic and alternative bond interactions. It is described by
the Hamiltonian:
\begin{equation}\label {fitrD}
   H = \sum_{i=1}^{N} \left((1-(-1)^{i}r)\sigma_{z}^{[i]} \sigma_{z}^{[i+1]} + D_{z}\;(\sigma_{x}^{[i]}
    \sigma_{y}^{[i+1]} - \sigma_{y}^{[i]} \sigma_{x}^{[i+1]})\right),
 \end{equation}
where $\sigma_{\alpha}^{[i]} \; (\alpha = x,y,z)$ are the Pauli
spin-$1/2$ operators at site $i$. The alternative bond interaction
is characterized by the relative strength $r$ of the exchange
coupling, and  $D_{z}$ is the $z$ direction component of the
Dzyaloshinskii-Moriya interaction arising from the spin-orbit
coupling. The  model is critical at $D_{z} \sim 0.263$, if  $r$ is
chosen to be $0.5$~\cite{Bo}.

The third model is quantum XYX model in an external magnetic field
described by the Hamiltonian:
\begin{equation}\label {fitxyx}
   H = \sum_{i=1}^{N} \left(\sigma_{x}^{[i]} \sigma_{x}^{[i+1]} + \Delta_{y}\sigma_{y}^{[i]}
    \sigma_{y}^{[i+1]} + \sigma_{z}^{[i]} \sigma_{z}^{[i+1]} + h\sigma_{z}^{[i]}\right),
 \end{equation}
where $\sigma_{\alpha}^{[i]} \; (\alpha = x,y,z)$ are the Pauli
operators at site $i$, $\Delta_{y}$ is a parameter describing the
rotational anisotropy, and $h$ is an external magnetic field. For
$\Delta_{y} = 0.25$, the critical magnetic field is
$h_{c}=3.206$~\cite{zhao,my,mc}.

\begin{figure}
\begin{overpic}[width=85mm,totalheight=50mm]{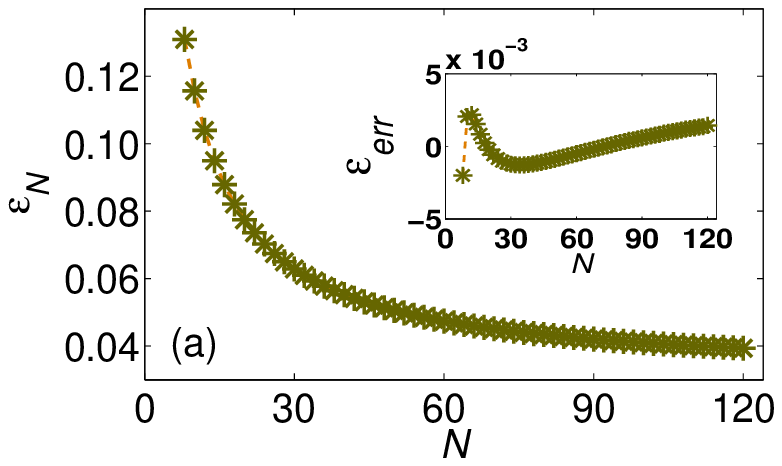}
\end{overpic}
\begin{overpic}[width=85mm,totalheight=50mm]{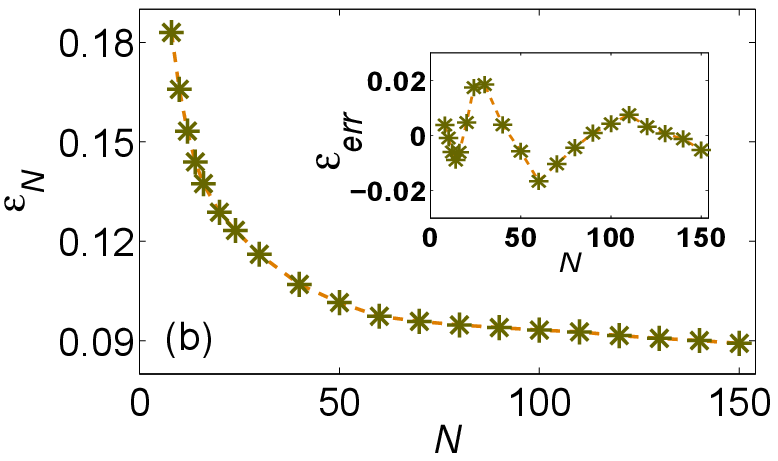}
\end{overpic}
\begin{overpic}[width=85mm,totalheight=50mm]{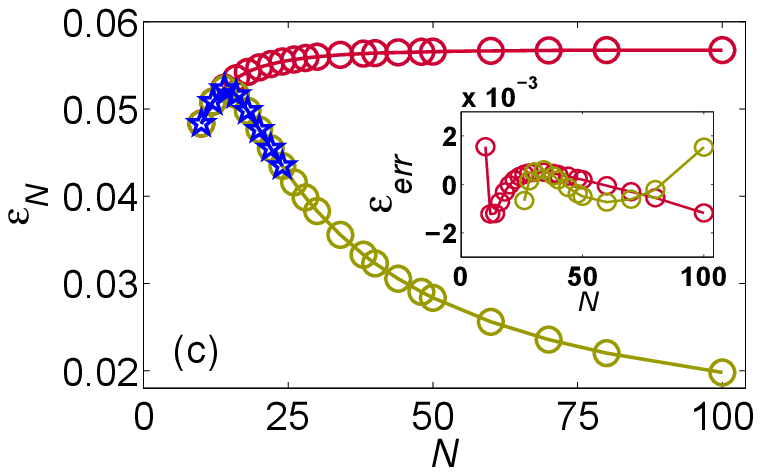}
\end{overpic}
\caption{(color online) Main: A scaling relation between the GE per
lattice site ${\mathcal E}_{N}$ and the chain size $N$, for quantum
Ising model (a), with the size $N$ ranging from 8 to 120, for
quantum Ising model with antisymmetric anisotropic and alternative
bond interactions (b), with the size $N$ ranging from 8 to 150, and
for quantum XYX model (c), with the size $N$ ranging from 10 to 100.
The data are fitted into ${\mathcal E}_{N}=a+b/N+f/N^{2}$, with
$a=0.030960$, $b=1.016076$ and $f=-1.712521$ in (a) and
$a=0.083320$, $b=0.966353$ and $f=-1.398155$ in (b). In (c), we have
$a=0.056892, b=0.001581$, and $f=-0.879631$, and $a=0.010193,
b=1.009600$ and $f=-5.005750$ , respectively, for the upper and
lower branches. Here, the upper (lower) branch is for the separable
states that are translation-invariant under one-site (two-site)
shifts, if the chain size is lager than a threshold. Indeed, the GE
per site is the same for two cases, if the chain size is less than
the threshold. In addition, the exact diagonalization results for
the GE per lattice site are displayed in five-pointed stars, up to
the chain size 24. The data match very well. Inset: The relative
fitting error $\varepsilon_{err}=(\varepsilon_{N}^{\rm
data}-\varepsilon_{N}^{\rm fit})/\varepsilon_{N}^{\rm fit}$ are less
than $2.1\times10^{-3}$ in (a), less than $1.8\times10^{-2}$ in (b),
and less than $1.5\times10^{-3}$ in (c), where $\mathcal{E}_N^{{\rm
fit}}$ is the value extracted from the fit and $\mathcal{E}_N^{{\rm
data}}$ is our simulation value for each $N$.}
  \label{figising}
   \end{figure}

{\it Simulation results.} For the transverse Ising model, we have
evaluated the GE per lattice site, as shown in
Fig.~\ref{figising}(a), with $b=1.016076$, consistent with our
previous result in Ref. ~\cite{Qian}. This yields the Affleck-Ludwig
$g$ factor $g=0.7032$. Compared to the exact value $g_{\rm
{fixed}}=1/\sqrt{2}$~\cite{affleck}, the relative error is
$|(g-g_{\rm {fixed}})|/g_{\rm {fixed}}=1.3 \times10^{-4}$. We
emphasize that, for this model, we have chosen the separable states
to be translation-invariant under one-site shifts, although there is
no a priori reason to argue that the GE per lattice site does not
depend on the unit cell of the separable states under translational
shifts. However, the same $g$ factor is yielded, even if the
separable states are translation-invariant under two-site shifts. In
fact, we have evaluated the GE per site without {\it any}
translation-invariant assumption, the same $g$ factor is achieved.
Thus, we conclude that {\it only} the smaller $g$ factor,
corresponding to the stable conformally invariant boundary
condition~\cite{CFT}, i.e., fixed boundary condition, is involved.

In Fig.~\ref{figising}(b), we plot the GE per lattice site,
$\mathcal{E}_N$, as a function of the number of the lattice sites,
$N$, for the spin-1/2 Ising chain with the coupling parameters
$r=0.5$ and $D_{z}=0.263$. The finite-size scaling for the GE per
lattice site follows ${\mathcal E}_{N}=a+b/N+f/N^{2}$, where the
coefficients are $a=0.083320$, $b=0.966353$ and $f=-1.398155$. It
yields the Affleck-Ludwig $g$ factor $g=0.715330$. Compared with the
exact value $g_{\rm fixed}= 1/\sqrt{2}$~\cite{affleck} for the
conformally invariant fixed boundary condition, the relative fitting
error is $|(g-g_{\rm fixed})|/g_{\rm fixed}=1.1 \times10^{-2}$. For
this model, we have chosen the separable states to be
translation-invariant under two-site shifts, given the alternative
bond interaction. However, we have evaluated the GE per site without
{\it any} translation-invariant assumption, the same $g$ factor is
achieved. Thus, we conclude that {\it only} the smaller $g$ factor,
corresponding to the stable conformally invariant boundary
condition~\cite{CFT}, i.e., fixed boundary condition, is involved.

In Fig.~\ref{figising}(c), we plot the GE per lattice site,
$\mathcal{E}_N$, as a function of the number of the lattice sites,
$N$, for quantum XYX model in an external field, with $\Delta_{y} =
0.25$, where the size $N$ is chosen from $10$ to $100$.  Remarkably,
it splits into two branches for a large enough chain size: one is
for the separable states that are translation-invariant under
one-site shifts, and the other is for the separable states that are
translation-invariant under two-site shifts. The exact
diagonalization for small-size chains up to 24 is also performed to
evaluate the GE per site without {\it any} translation-invariant
assumption. However, it only reproduces the second branch, as seen
in Fig.~\ref{figising}(c). A proper explanation for this unexpected
result is that, for this model, the translational invariance under
one-site shifts constitutes a true constraint to separable states.
That is, the closest separable state with the one-site unit cell is
different from the closest separable state with the two-site unit
cell. Thus the maximum fidelity $\Lambda_{{\rm max}}$ is larger for
the latter, implying that the GE per lattice site is smaller, as
long as the chain size is large enough. This explains the presence
of a threshold in the chain size: the GE per site is the same for
both the one-site and two-site unit cells, if the size is less than
the threshold. Two sets of the data are separately fitted into the
scaling function ${\mathcal E}_{N}=a+b/N+f/N^{2}$, with the
coefficients $a=0.056892, b=0.001581$, and $f=-0.879631$, and
$a=0.010193, b=1.009600$ and $f=-5.005750$, respectively, yielding
$g_{\rm XYX}=0.9994$ and $g_{\rm XYX}=0.7048$. They are consistent
with the exact values $g_{\rm free}=1$ and $g_{\rm fixed}=
1/\sqrt{2}$~\cite{affleck} for conformally invariant free and fixed
boundary conditions in the Ising universality class.

{\it Summary.} In this paper, we have developed a scheme to
efficiently compute the GE per lattice site for quantum many-body
spin systems on a periodic finite-size chain in the context of a
tensor network algorithm based on the matrix product state
representations. The computational cost does not depend on the chain
size. A systematic test is performed for three prototypical critical
quantum spin chains belonging to the same Ising universality class.
The simulation results lend strong support to the previous claim
that the leading finite-size correction to the GE per lattice site
is universal~\cite{Qian}, with its remarkable connection to the
celebrated Affleck-Ludwig boundary entropy corresponding to a
conformally invariant boundary condition~\cite{Misguich}. For all
the models tested, the simulated $g$ is compared to the exact $g$
factor from conformal field theory, with the relative error less
than $1.1\times10^{-2}$. It appears that the boundary entropy
corresponding to the smallest $g$ factor is always involved. This is
somewhat expected, since this $g$ factor characterizes a stable
fixed point along a boundary renormalization group flow, according
to the Affleck-Ludwig $g$ theorem~\cite{affleck}. Remarkably, for
quantum XYX model in an external field, either conformally invariant
free or fixed boundary condition appears, depending on the one- or
two-site unit cell of the translation-invariant separable states.

{\it Acknowledgements.} We thank J.O. Fj{\ae}restad and R. Or\'{u}s
for their inspiring discussions. The work is partially supported by
the National Natural Science Foundation of China (Grant No:
10874252). XJL and JHL are supported by the Fundamental Research
Funds for the Central Universities (Project No. CDJXS11102213) and
by Chongqing University Postgraduates' Science and Innovation Fund
(Project No.: 200911C1A0060322).


\end{document}